\documentclass[fleqn,usenatbib]{mnras}

\usepackage{newtxtext,newtxmath}

\usepackage[T1]{fontenc}

\DeclareRobustCommand{\VAN}[3]{#2}
\let\VANthebibliography\thebibliography
\def\thebibliography{\DeclareRobustCommand{\VAN}[3]{##3}\VANthebibliography}

\usepackage{graphicx}

\usepackage[T1]{fontenc}
\usepackage[utf8]{inputenc}
\usepackage{amsmath}
\usepackage{color}
\usepackage{mathtools}
\usepackage{hyperref}
\usepackage{natbib}

\newcommand{\dd}[1]{\mathrm{d}#1 \,}

\newcommand{\pdev}[2]{\frac{\partial #1}{\partial #2}}
\newcommand{\pdevn}[3]{\frac{\partial^{#3} #1}{\partial #2^{#3}}}

\newcommand{\dev}[2]{\frac{\mathrm{d}#1}{\mathrm{d}#2}}
\renewcommand{\v}[1]{\boldsymbol{#1}}

\newcommand{\crl}[1]{\langle #1 \rangle}

\newcommand{\addcom}[1]{}

\title[CR confinement by micromirrors]{Cosmic-ray confinement in radio bubbles by micromirrors}

\author[R. J. Ewart et al.]{Robert J. Ewart$^{1,2}$\thanks{E-mail: robert.ewart@physics.ox.ac.uk (KTS)}, Patrick Reichherzer$^{1,3}$, Archie F. A. Bott$^{1,4}$, Matthew W. Kunz$^{5,6}$,\newauthor and Alexander A. Schekochihin$^{1,7}$
\\
$^{1}$Department of Physics, University of Oxford, Oxford, OX1 3PU, UK\\
$^{2}$Balliol College, Oxford, OX1 3BJ, UK\\
$^{3}$Exeter College, Oxford, OX1 3DP, UK\\
$^{4}$Trinity College, Oxford, OX1 3BH, UK\\
$^{5}$Department of Astrophysical Sciences, Princeton University, Peyton Hall, Princeton NJ 08544, USA\\
$^{6}$Princeton Plasma Physics Laboratory, Princeton University, PO Box 451, Princeton NJ 08543, USA\\
$^{7}$Merton College, Oxford OX1 4JD, UK
}

\date{Accepted XXX. Received YYY; in original form ZZZ}

\pubyear{2024}

\begin{document}
\label{firstpage}
\pagerange{\pageref{firstpage}--\pageref{lastpage}}
\maketitle

\begin{abstract}
Radio bubbles, ubiquitous features of the intracluster medium around active galactic nuclei, are known to rise buoyantly for multiple scale heights through the intracluster medium (ICM). It is an open question how the bubbles can retain their high-energy cosmic-ray content over such distances. We propose that the enhanced scattering of cosmic rays due to micromirrors generated in the ICM, as proposed recently by Reichherzer et al. (2023), is a viable mechanism for confining the cosmic rays within bubbles and can qualitatively reproduce their morphology. We discuss the observational implications of such a model of cosmic-ray confinement.
\end{abstract}



\section{Introduction}
Amongst the various morphologies exhibited in radio emission by active galaxies, one of the most prominent is the so-called radio `bubble' (\citealt{Fanaroff_1974,Churazov_2000,Churazov_2001,de_Gasperin_2012,Velovic_2023}), christened as such by \cite{Gull_1973} for their analogy to bubbles of air rising through liquid. The brightness of these structures in radio (between several tens of MHz and a few GHz, \citealt*{Dunn_2005}) being roughly consistent with an absence of X-rays (see, e.g., \citealt{Bohringer_1993,Birzan_2004,McNamara_2007}) indicates that they are filled with high-energy particles emitting synchrotron radiation in the presence of a magnetic field. These bubbles are observed at a range of distances from the centres of cool-core clusters, where they are sourced by Active Galactic Nuclei (AGN), implying there must be some mechanism, presumably magnetic in nature, by which high-energy cosmic rays (CRs) are confined to within the bubble for the duration of the bubble's rise. Recently \cite{Reichherzer_2023} have proposed that, within the ICM, the predominant scattering of sub-$\mathrm{TeV}$ CRs is due to deflections by small-scale `micromirrors', which should arise naturally, generated by the mirror instability, in high-$\beta$ plasmas such as the ICM (\citealt{Schekochihin_2005}; \citealt*{Kunz_2022}). In this paper, we show that such scattering is capable of efficiently confining CRs within bubbles, qualitatively mimicking the sharp boundary seen in radio emission.

The life of a typical bubble begins with the inflation of a radio lobe by the jet of an AGN. After initial supersonic expansion, the lobe is further slowly inflated in pressure balance with the ICM into which it is injected. By virtue of its underdensity with respect to the surrounding matter, the lobe begins to rise, eventually detaching from the AGN and becoming a bubble. Throughout the bubble's rise, it is deformed under the influence of hydrodynamic and magnetohydrodynamic effects (\citealt{Churazov_2001,Bruggen_2003}). Such evolution has been studied extensively not only for its own sake, but also because the bubbles are considered to be a potent source of energy that can account for a significant portion of the heating required to obviate the cooling-flow problem and regulate AGN feedback (see, e.g., \citealt{Churazov_2000, Peterson_2006, McNamara_2007, Werner_2018}). The exact method by which this energy is deposited into the ICM has been the subject of numerous studies, both analytical and numerical (\citealt{Churazov_2001}; \citealt*{Reynolds_2015}; \citealt{Bambic_2019}; \citealt*{Chen_2019}).

Despite extensive studies of the various mechanisms by which the bubble deposits energy into the ICM, there is still some debate as to which mechanisms protect the bubbles from a host of hydrodynamic instabilities (see, e.g., \citealt{Ruszkowski_2007,Guo_Mathews_2012b,Kingsland_2019}). Putting aside the question of the longevity of the bubbles themselves, the sharp boundary of the bubble in radio and the lack of observed $\gamma$-ray emission in galaxy clusters (see, e.g., \citealt{Ahnen_2016, Prokhorov_2017}), suggests that CRs are well confined within the bubble for long times. How they manage that is a question that remains open. Previous models for CR confinement in bubbles have relied on the large-scale magnetic-field structure around the bubble, viz., draping of magnetic fields over the bubble (see, e.g., \citealt{Dursi_Pfrommer_2008}), or on the internal structure of the magnetic field, which are thought to confine CRs by exploiting the lack of field lines crossing the bubble boundary. However, such models may be vulnerable to the objection that any field lines that do connect the bubble's interior to its exterior, e.g., those that could be created by reconnection on the bubble's surface, would become a highway along which the CRs could escape\footnote{Indeed, the extent to which the uniformity of the draped field line influences CR confinement within inflated jets was investigated by \cite{Desiati_Zweibel_2014}, concluding non-uniformities significantly boost diffusion across field lines, in the context of the much closer, and younger, Fermi bubbles \citep{Yang_2012}.}. Even in the absence of such violent escape of CRs, streaming along magnetic-field lines would naturally lead to significant variation in the extent of radio emission as a function of observed frequency---simply put, older CRs would have travelled further along magnetic-field structures creating extended emission at lower frequencies. This is precisely what is not seen in observations of radio bubbles (see, e.g., \citealt{de_Gasperin_2012,Brienza_2021}). We therefore propose that it is CR scattering off small-scale fluctuations in the weak magnetic field outside the bubble that, paradoxically, can serve as a good mechanism of confinement. 

Intuitively the explanation for this form of confinement is as follows. CRs leaving the bubble in the direction of the bubble's motion find themselves scattering off fluctuations in the magnetic field, and becoming diffusive after a single mean free path. Like a ballistic tortoise racing against a diffusive hare, the bubble will continue its buoyant motion at roughly constant velocity, which overtakes the diffusive motion of the recently escaped CRs over sufficiently long times. This allows the bubble effectively to recapture some of the escaping CRs, maintaining the sharp boundary seen in observations (see, e.g., \citealt{de_Gasperin_2012, Brienza_2021}). In the rest frame of the bubble, this amounts to the CRs scattering in the flow sweeping past the bubble and thus being driven back onto the bubble. Underneath the bubble (i.e., in the opposite direction to the bubble velocity), the same diffusive behaviour applies, but CRs clearly cannot be swept back up. However, provided the bubble creates a sufficiently extended wake behind it (as indeed seen in numerical simulations, e.g., by \citealt*{Zhang_2018}; \citealt{Zhang_2022}), the flow immediately behind the bubble will also be rising, rendering the loss of CRs behind the bubble entirely diffusive. 

Clearly, any diffusive process is necessarily a leaky and imperfect confinement mechanism. It must be the case that the CRs with larger diffusion coefficients escape from the bubble more easily. This can be quantified by the following, largely dimensional, argument (\citealt{Mathews_2007}). A CR escaping the bubble with a diffusion coefficient $\kappa$ will, on average, travel a distance $\sim \sqrt{\kappa t}$ in time $t$. Even if the CR diffuses entirely vertically, the bubble, moving at speed $u_{b}$, will overtake the CR after a time $t\sim \kappa/ u_{b}^{2}$. Therefore, for the CR to escape the bubble rather than be recaptured, it must diffuse out of the path of the bubble---a horizontal distance comparable to the radius~$r_{b}$ of the bubble---in a time $t \sim \kappa/ u_{b}^{2}$. Hence, for CRs to be recaptured by the bubble upon their escape, their diffusivity must be sufficiently small so that
\begin{equation}
\label{eqn:S0:E1}
\frac{\kappa}{u_{b}r_{b}} \lesssim 1.
\end{equation} 
We will find that the conventionally estimated diffusivity of CRs, due to scattering by the streaming instability, is insufficient to satisfy (\ref{eqn:S0:E1}) \citep{Subedi_2017,Krumholz_2020}. In contrast, the scattering mechanism proposed recently by \cite{Reichherzer_2023} provides a sufficiently low diffusion coefficient to confine CRs below $1\,\mathrm{TeV}$ and reproduce the observed radio-bubble morphology. 

The rest of this paper is structured as follows. In section \ref{Section:micromirror_diffusion_coefficient}, we discuss the micromirror scattering mechanism for CRs below $1\,\mathrm{TeV}$ based on \cite{Reichherzer_2023}, and explain why the micromirrors should be endemic to the vicinity of the bubble. In section \ref{Section:central_concept}, we show analytically and numerically that, even when the scattering centres are exclusively outside the bubble, the effect of the scattering can create a sharp bubble boundary with CRs confined either inside the bubble or within a thin layer of the ICM around it. Using the energy dependence of the scattering frequency, we obtain a prediction of how the thickness of this layer of CRs around the bubble should depend on particle energy. Furthermore, we show that the energy dependence of the scattering frequency gives rise to an interesting effusive behaviour, making the CR energy spectrum outside the bubble distinct from (``harder'' than) that inside it. Finally, in section \ref{Section:Conclusion}, we summarise the progress that has been made and discuss the role of other sources of CR confinement.
\section{Diffusivity due to micromirrors}
\label{Section:micromirror_diffusion_coefficient}
While typical estimates of the speed of bubble's buoyant rise through the ICM are on the order of the Keplerian velocity around the cluster centre (see, e.g., \citealt{Churazov_2001}), the CRs propagate at a velocity very close to the speed of light. In the absence of any forces, therefore, they would leave the bubble in a single light-crossing time. In reality, the bubble is permeated and surrounded by magnetic fields~$\v{B}$. These magnetic fields shape the velocity $\v{v}$ of the CRs, forcing them to follow field lines in accordance with the Lorentz-force law, viz.,
\begin{equation}
\label{eqn:S1:E1}
\dev{\v{v}}{t} = \frac{q}{m \gamma}\frac{\v{v}\times \v{B}}{c},
\end{equation}
where $q$ and $m$ are the charge and rest mass of the CR, respectively, and $\gamma = (1-|\v{v}|^{2}/c^{2})^{-1/2}$ is its Lorentz factor. 

The confinement of field-line-following CRs can only be efficient if no field lines cross the bubble boundary. However, one might expect that reconnection on the bubble surface would quickly engender such stray field lines, reducing the effectiveness of confinement. In such cases, there must be some further effect limiting the escape of CRs along field lines. For example, a CR following the large-scale magnetic field will pitch-angle scatter off any small-scale (meaning smaller than the CR's gyroradius) fluctuating magnetic fields~$\delta \v{B}$, and thus migrate onto a different field line. After many such scatterings, its direction can entirely reverse. Following \cite{Reichherzer_2023}, we can use (\ref{eqn:S1:E1}) to estimate the amount by which such a CR will be scattered by fluctuations of amplitude $\delta B$ in a time $\delta t$:
\begin{equation}
\label{eqn:S1:E2}
\frac{|\delta \v{v}|}{c} \sim \frac{q\delta B}{\gamma m c^{2}}c \delta t \sim \frac{c \delta t}{r_{\mathrm{g}}}\frac{\delta B}{B},
\end{equation}
where $r_{\mathrm{g}} = \gamma m c^{2}/ qB$ is the CR gyroradius. The CR will spend a time $\delta t \sim l/c$ traversing a magnetic fluctuation of typical scale $l$. If the ICM is assumed to be infested with such structures and the deflections from each structure add up like a random walk, then (\ref{eqn:S1:E2}) can be used to estimate the scattering frequency---the rate at which the pitch angle of a CR is changed by an order-unity amount---as 
\begin{equation}
\label{eqn:S1:E3}
\nu_{\mathrm{mm}}\sim \frac{|\delta \v{v}|^{2}}{c^{2}\delta t} \sim \frac{c l}{r_{\mathrm{g}}^{2}}\left( \frac{\delta B}{B}\right)^{2}.
\end{equation} 

This formula makes several features obvious. First, the scattering frequency strongly increases with decreasing CR gyroradius (or, equivalently, the CR energy, since the two are linearly proportional to each other for relativistic CRs). Secondly, and intuitively, the larger is the relative amplitude of the fluctuating magnetic field, the more efficient will be the scattering of CRs. As pointed out by \cite{Reichherzer_2023}, in the ICM plasma, there is a myriad of ways of creating microscale magnetic fluctuations by kinetic instabilities enabled in the high-$\beta$ regime (\citealt{Schekochihin_2006}; \citealt*{Bott_2023}). These kinetic instabilities arise because, in the high-$\beta$ regime, low-Mach number turbulent motions are able to cause changes in the magnitude of the magnetic field, which in turn drive pressure anisotropies that compete with magnetic forces, and can excite ion-Larmor-scale kinetic instabilities. Amongst this family of instabilities, the mirror instability, generated in regions of increasing magnetic-field strength, is special in that it saturates with $\delta B/B\sim 1/3$ (\citealt*{Kunz_2014}; \citealt*{Riquelme_2015}; \citealt{Melville_2016}), making it a prime candidate for scattering cosmic rays. Making a rough estimate of the relevant ICM parameters, one finds the CR diffusion coefficient due to micromirrors to be \citep{Reichherzer_2023}
\begin{equation}
\label{eqn:S1:E4}
\kappa \sim 10^{30}\left(\frac{E}{\mathrm{TeV}} \right)^{2} \mathrm{cm}^{2}\mathrm{s}^{-1},
\end{equation}
where $E$ is the CR energy. Here the key step is to compare it to the dimensional estimate~(\ref{eqn:S0:E1}). Noting the typical values of the bubble radius~$r_{b} \sim 5-10\, \mathrm{kpc}$ and velocity $ u_{b} \sim 100-400\,\mathrm{km}\,\mathrm{s^{-1}}$ (e.g., \citealt{Zhang_2018}), one finds that
\begin{equation}
\label{eqn:S1:E5}
\frac{\kappa}{u_{b}r_{b}}\sim \left(\frac{E}{\mathrm{TeV}} \right)^{2}\left(\frac{u_{b}}{200\,\mathrm{km}\,\mathrm{s^{-1}}} \right)^{-1}\left(\frac{r_{b}}{10\,\mathrm{kpc}} \right)^{-1}.
\end{equation}
This implies that, for energies much larger than a $\mathrm{TeV}$, the diffusion coefficient is most likely too large to be of much assistance with CR confinement in bubbles. In contrast, for CR energies lower than a few $100\,\mathrm{GeV}$, micromirror diffusion could indeed be consequential. While the precise value of the diffusion coefficient is subject to a number of factors (we refer the reader to \citealt{Reichherzer_2023} for details), it should be noted that the region of compressing (draped) magnetic field around the bubble, of characteristic size $\sim r_{b}$ (see, e.g., \citealt{Dursi_Pfrommer_2008}), creates the perfect environment for micromirrors to thrive because, as discussed above, mirrors are driven by positive pressure anisotropy, which is created in regions where the magnetic field's strength is increasing.

One could still object that it is unclear how small one requires the ratio~(\ref{eqn:S1:E5}) to be for good confinement, or indeed whether some power of the dimensionless factor of $c/u_{b}$ should appear for some unknown reason. We must, therefore, scope out whether our picture of the bubbles can hold water, or indeed CRs. This is done, both numerically and analytically, in the next section.
\section{Proof of concept and mock simulation}
\label{Section:central_concept}
\begin{figure}
\centering
\includegraphics[width = 1.0\linewidth]{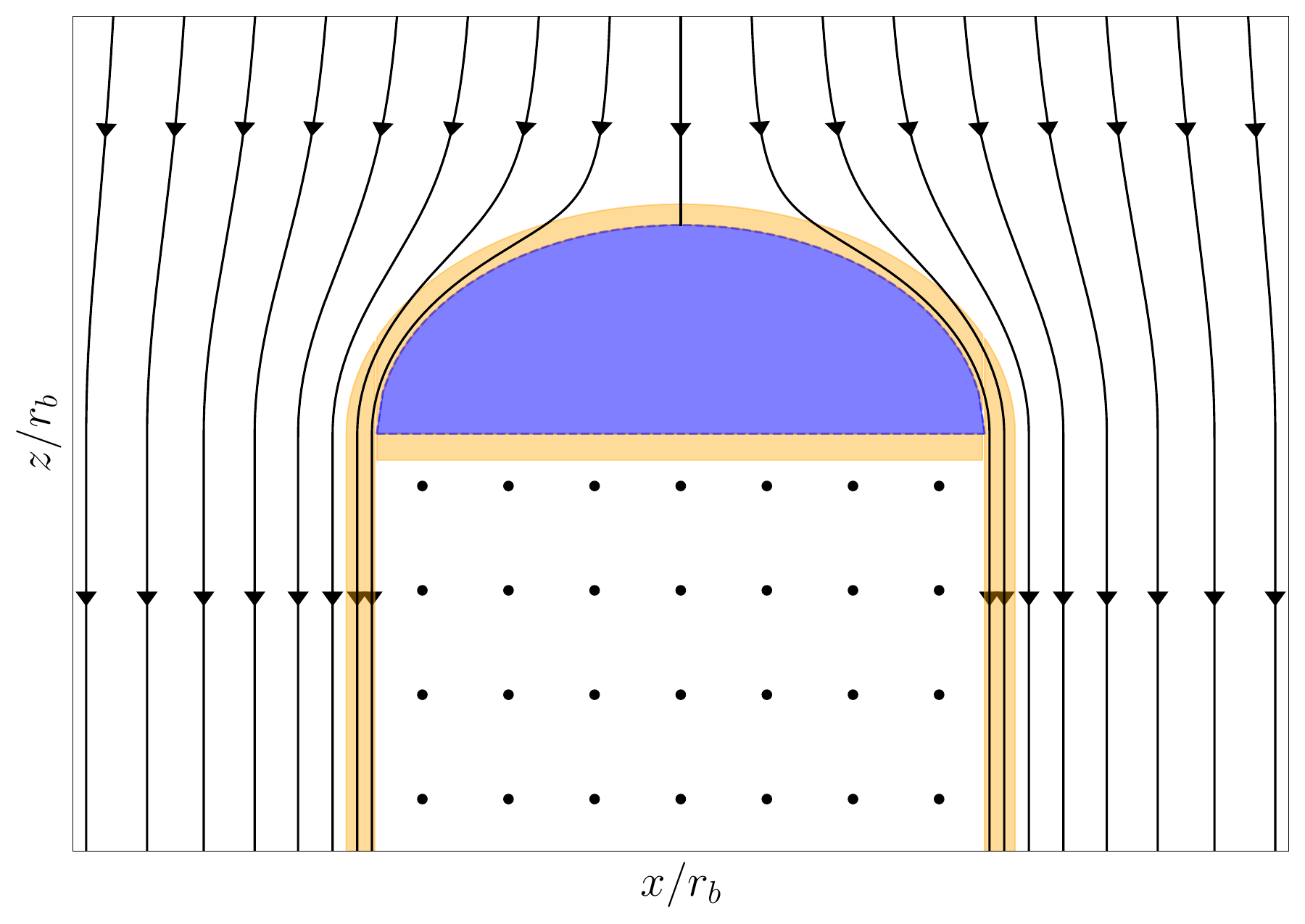}
\caption{A schematic of a 2D slice of the fluid flow around the bubble used for our mock simulation. The blue shaded region represents the nominal interior of the bubble, in which the CRs propagate ballistically. The expected area in which CRs will diffuse off the bubble is shaded in orange. The arrowed black lines are the streamlines of the flow around the bubble. Beneath the bubble, the flow is stationary, crudely approximating a wake.}
\label{Fig:Schematic}
\end{figure}
We first investigate our model of confinement numerically. Clearly, a full solution of the problem would require an accurate magnetohydrodynamic model of the nearly collisionless ICM coupled to a kinetic or fluid model of CRs (see, e.g., \citealt{Zweibel_2017}; \citealt*{Weber_2022}). However, the choice of a fluid magnetohydrodynamic model in a nearly collisionless environment is still an open area of research (see, e.g., \citealt{Squire_2019,Squire_2023,Kunz_2022}), as is the fluid dynamics of CRs. Furthermore, many studies of the buoyant rise of bubbles are plagued by hydrodynamic instabilities that shred the bubbles long before they have risen to the distances at which they are observed (e.g., \citealt{Dong_2009,Reynolds_2015}). We therefore take an extremely simplified toy model in which the flow is externally prescribed. Such an approach may be partially justified \textit{a posteriori}, should we find that the CRs are well confined within the bubble (which we will), since there is then at least no inconsistency with the bubble having a boundary.

The details of the numerical simulation are presented in Appendix~\ref{App:Numerical_details}. The basic concept of it is as follows. Within the bubble, the CRs move ballistically as if under no force. This is clearly a simplification of the exact, field-line following trajectory that the CRs will take. Outside the bubble, the CRs scatter off micromirrors, but crucially, they do so in the rest frame of the flow $\v{u}$ that they encounter, because mirror-unstable fluctuations have no phase velocity relative to the bulk plasma. We take the imposed flow to be a 3D, incompressible, irrotational flow above a half-sphere, while below the sphere, to simulate the effect of a wake, the flow is taken to be entirely vertical. A schematic of this flow is shown in Figure~\ref{Fig:Schematic}, the explicit expressions for it are given in Appendix~\ref{App:Numerical_details}. Clearly, the wake of this bubble is a major simplification, since it effectively implies that the bubble causes a rigid column of fluid to rise behind it. Simulations with rigid bubbles have exhibited wakes that extend over several tens of kiloparsecs entraining a large amount of material, albeit via vortices, in the wake of the bubble (e.g., \citealt{Zhang_2022}). We therefore do not hope to capture accurately the density of CRs within the wake. Rather we include the rising column to ensure that there is no spurious flow beneath the bubble capable of stealing CRs directly from it. This toy model will prove sufficient to illustrate the physics relevant to our proposed theory of CR confinement.
\begin{figure*}
\centering
\includegraphics[scale=0.5]{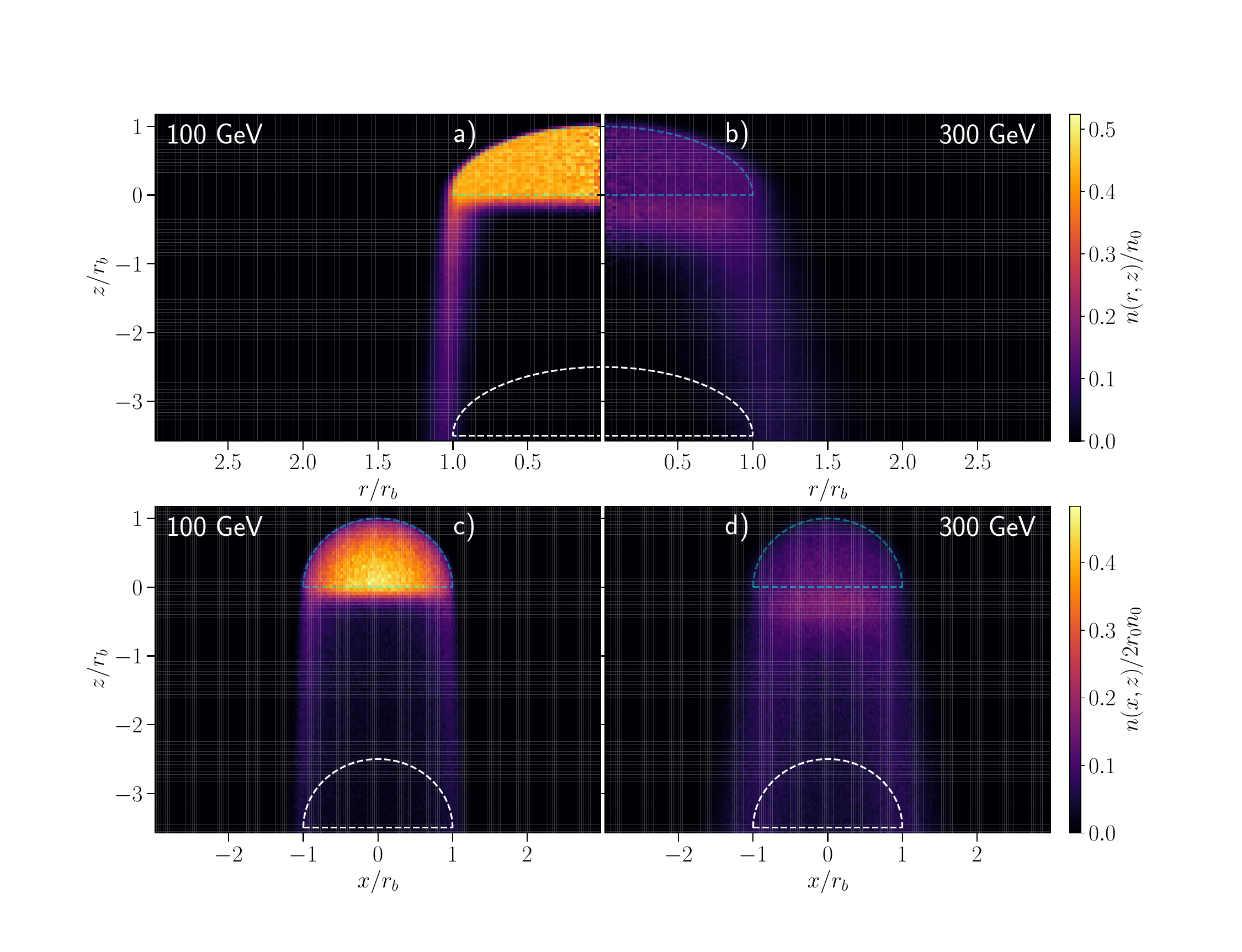}
\caption{The density of CRs attained numerically with populations at distinct energies ($100\,\mathrm{GeV}$ and $300\, \mathrm{GeV}$), and, therefore, diffusion coefficients, after the bubble rose a distance of $3.5 r_{b}$. Both bubbles were chosen to have a radius of $10\, \mathrm{kpc}$ and to rise at constant velocity of $200\, \mathrm{km}\,\mathrm{s}^{-1}$. Panels (a) \& (b) show the azimuthally averaged densities, panels (c) and (d) show the same densities in projection. The densities are normalised to the maximum value that the initial density (all CRs inside the bubble) could take in that projection. Dashed lines indicate the initial (white) and final (blue) positions of the nominal boundary of the bubble.}
\label{Fig:numerical result}
\end{figure*}

Our central result---that scattering by micromirrors efficiently confines CRs of low energy---is illustrated in Figure~\ref{Fig:numerical result}, which shows the density (normalised to the initial density $n_{0} = 3N/2\pi r_{b}^{3}$ of $N$ CRs in a hemisphere of radius~$r_{b}$) of two populations of CRs initialised at two different energies, after the bubble has risen to a height equal to $3.5$ times its radius, which was taken to be $10\, \mathrm{kpc}$ for these simulations. The CRs at $100$ GeV are confined remarkably well to the bubble, while the CRs at $300$ GeV have leaked out of the bubble left, right, centre, and, predominantly, downwards. As anticipated, we see that the upstream bubble boundary is much thinner than the bubble for both energies, because the CRs are being swept back onto the bubble. 

In what concerns the situation beneath the bubble, Figure \ref{Fig:numerical result} provides an immediate refinement to our picture of how bubbles leak CRs. All leakage occurs via the sides of the bubble as the CRs that diffuse horizontally are swept away with the flow (see the shaded orange region of Figure \ref{Fig:Schematic}). Clearly, for lower-energy CRs, one can estimate the rate of loss from the volume of ICM containing CRs that is swept away from bubble. This statement can be made more quantitative in the following way. Around the bubble there is a thin layer, of width $\Delta r$, containing CRs embedded in the ICM flow. In a short time $\Delta t$, the bubble sheds a cylindrical shell of height $\sim u_{b}\Delta t$ of this layer due to the flow dragging it away. If the CR density in the bubble and layer is roughly~$n$, then the number of CRs lost in this short time is of order $2\pi r_{b}n u_{b}\Delta t\Delta r $. Since the lost CRs come from the bubble, this gives us an estimate for the time dependence of the bubble density:
\begin{equation}
\label{eqn:S1:E5.5}
\frac{2\pi}{3}r_{b}^{3}\dev{n}{t} \sim -2\pi r_{b}n u_{b}\Delta r \implies \dev{n}{t} \sim -\frac{u_{b}\Delta r}{r_{b}^{2}}n.
\end{equation} 
Thus, we see that the rate of the CR loss depends crucially on the width $\Delta r$ of the layer of CRs draped over the bubble. To determine this width, we turn towards a toy analytical model for the diffusion of CRs around the bubble.
\subsection{The Blasius boundary layer and CR confinement time}
\label{subsection:Blasius}
\begin{figure}
\centering
\includegraphics[width = \linewidth]{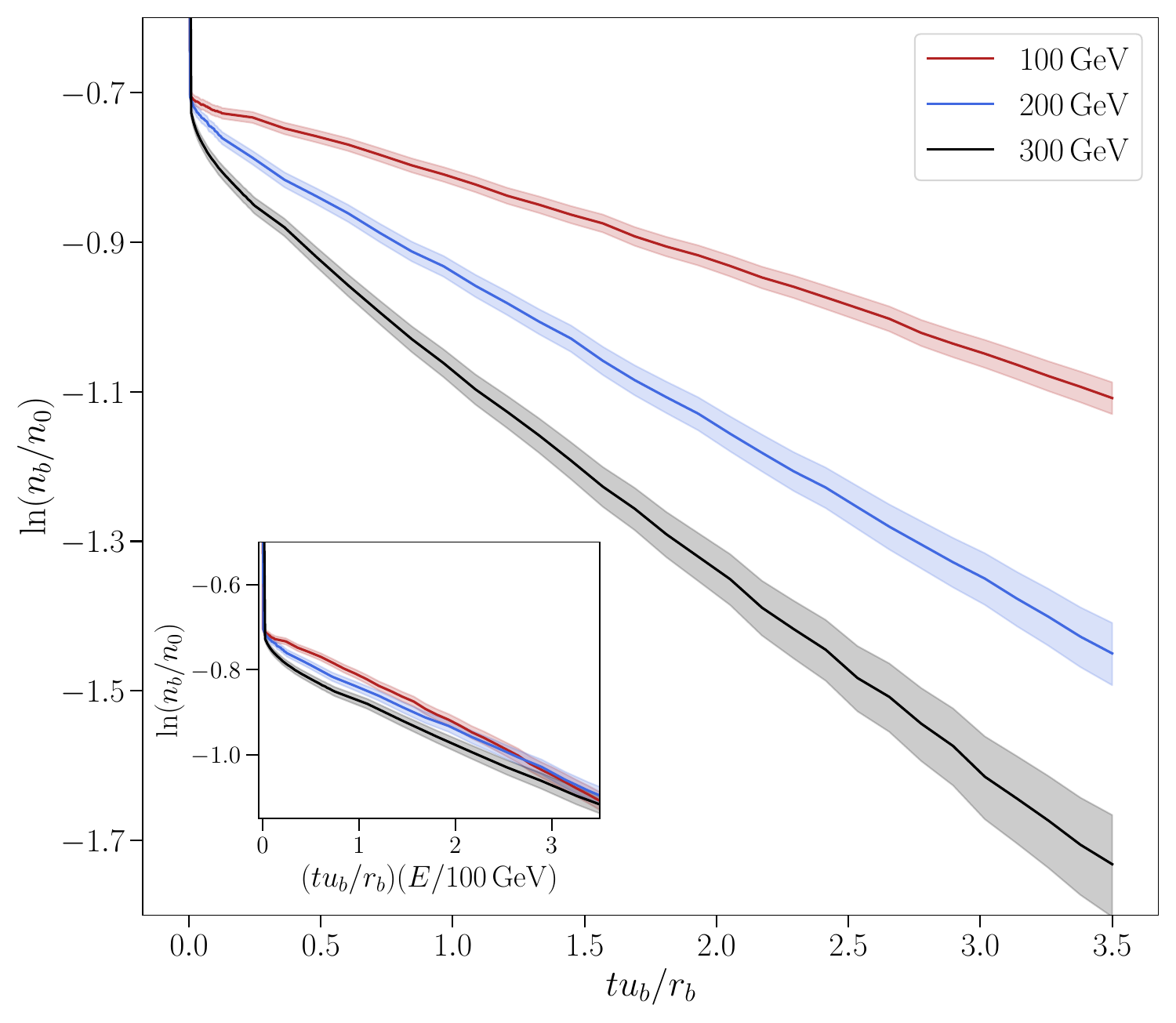}
\caption{The time evolution of the CR density within bubbles (with shading showing error bars of two standard deviations) initialised with CRs at three different energies,~$100\,\mathrm{GeV}$, $200\,\mathrm{GeV}$, and $300\,\mathrm{GeV}$. Inset shows the same evolution with the time axis scaled by the initial CR energy, which is the relevant variable in our order-of-magnitude estimate (\ref{eqn:S1:E7}).}
\label{Fig:BubbleDensity}
\end{figure}
As discussed above, the correct model for the hydrodynamics of CRs is an open area of research. However, for our toy model, we again take the simplest possible approximation for the evolution of the CR density outside the bubble. The CRs of any given energy diffuse with a diffusion coefficient $\kappa$ and are advected by the velocity field~$\v{u}$. Thus, outside the bubble, a model equation for the CR density $n$ is 
\begin{equation}
\label{eqn:S1:E6}
\pdev{n}{t} + \v{u}\v{\cdot}\nabla n = \kappa \nabla^{2}n.
\end{equation} 
Of course, this equation neglects a number of things.

First, it neglects any changes in the CR energy, and hence the resulting changes in their diffusion coefficient, due to acceleration mechanisms left outside our model (such as synchrotron losses), or indeed due to \cite{Fermi_1949} acceleration from scattering in the rest frame of the flow, which is present in our model. We may justify the neglect of synchrotron losses by noting that the typical synchrotron loss time is~{$t_{\mathrm{sync}} \sim 10^{7}(B/\mathrm{\mu G})^{-2} (E/\mathrm{TeV})^{-1}\, \mathrm{yr} $} (see, e.g., \citealt{Jackson_1998,Hess_2016}) which, for CRs with energies much lower than a $\mathrm{TeV}$ is substantially longer than the bubble rise time. As for Fermi acceleration, since the flow speed is a small fraction of the speed of light, we can expect the acceleration to be relatively weak over the timescales that we are considering (we justify this \textit{a posteriori} in Appendix \ref{App:Energy}). 

Secondly, the CR density within the bubble itself is not specified by~(\ref{eqn:S1:E6}). Thankfully, the light-crossing time~$r_{b}/c$, which is the timescale on which density perturbations will be ironed out inside the bubble, is much shorter than the bubble rise time~$r_{b}/u_{b}$. As a result, the CR density in the bubble interior only functions as a boundary condition fixing the density at the bubble-ICM interface. This boundary condition will evolve in time as the density within the bubble decreases, but, provided the confinement is good, this boundary condition will evolve slowly allowing the density of CRs outside the bubble to reach a quasi-steady profile, with an amplitude set by their density at the boundary.

The solution of the advection-diffusion equation~(\ref{eqn:S1:E6}) near a boundary is a well-studied topic (see, e.g., \citealt{Batchelor_1967,Landau_1987}). In the ICM above the bubble, this problem is mathematically identical to the problem of the width of cold fronts due to thermal conduction (studied, for instance, by \citealt{Churazov_2004,Xiang_2007}) with CR density taking the place of temperature. According to \cite{Xiang_2007}, a boundary layer of thickness $\Delta r$ given by
\begin{equation}
\label{eqn:S1:E6.5}
\frac{\Delta r}{r_{b}} \sim \sqrt{\frac{\kappa}{u_{b}r_{b}}}
\end{equation}
forms around the surface of the bubble. This is obvious if one notes that, in a steady state, the diffusive term in (\ref{eqn:S1:E6}) pushing CRs away from the bubble must balance the advective term sweeping them back onto the bubble; since the diffusive term scales as $\kappa n/\Delta r^{2}$ and the advective term scales as $u_{b}n/r_{b}$, one immediately arrives at~(\ref{eqn:S1:E6.5}). The full solution and scaling analysis of (\ref{eqn:S1:E6}), extended to the entire region~$0\leq\theta\leq \pi/2$, are given in Appendix \ref{App:Asym}. Combining the estimate~(\ref{eqn:S1:E6.5}) for the width of the layer with~(\ref{eqn:S1:E5}) for the CR diffusion coefficient upgrades the estimate (\ref{eqn:S1:E5.5}) of the density depletion rate to
\begin{equation}
\label{eqn:S1:E7}
\frac{1}{n}\dev{n}{t} \sim -\frac{u_{b}}{r_{b}}\sqrt{\frac{\kappa(E)}{u_{b}r_{b}}},
\end{equation}
giving a typical confinement time
\begin{equation}
\label{eqn:S1:E7p5}
t_{\mathrm{conf}} \sim \left(\frac{E}{\mathrm{TeV}} \right)^{-1}\left(\frac{u_{b}}{200\,\mathrm{km}\,\mathrm{s^{-1}}} \right)^{-1/2}\left(\frac{r_{b}}{10\,\mathrm{kpc}} \right)^{3/2} 100\, \mathrm{Myr}.
\end{equation}
A comparison with the results of numerical simulation in Figure \ref{Fig:BubbleDensity} shows good agreement with the general trend.

This result shows that the use of the micromirror model of \cite{Reichherzer_2023} is essential. As discussed in \cite{Reichherzer_2023}, CR diffusion due to the more conventional scattering mechanisms (turbulence, streaming instability, etc.) is comparable to the diffusion due to micromirrors for $\sim 1\,\mathrm{TeV}$ CRs. From (\ref{eqn:S1:E7}) and Figure \ref{Fig:BubbleDensity}, we see, therefore, that those conventional scattering schemes would be incapable of confining CRs efficiently. 
\subsection{Hardening of CR spectrum outside bubbles}
\begin{figure}
\centering
\includegraphics[width=\linewidth]{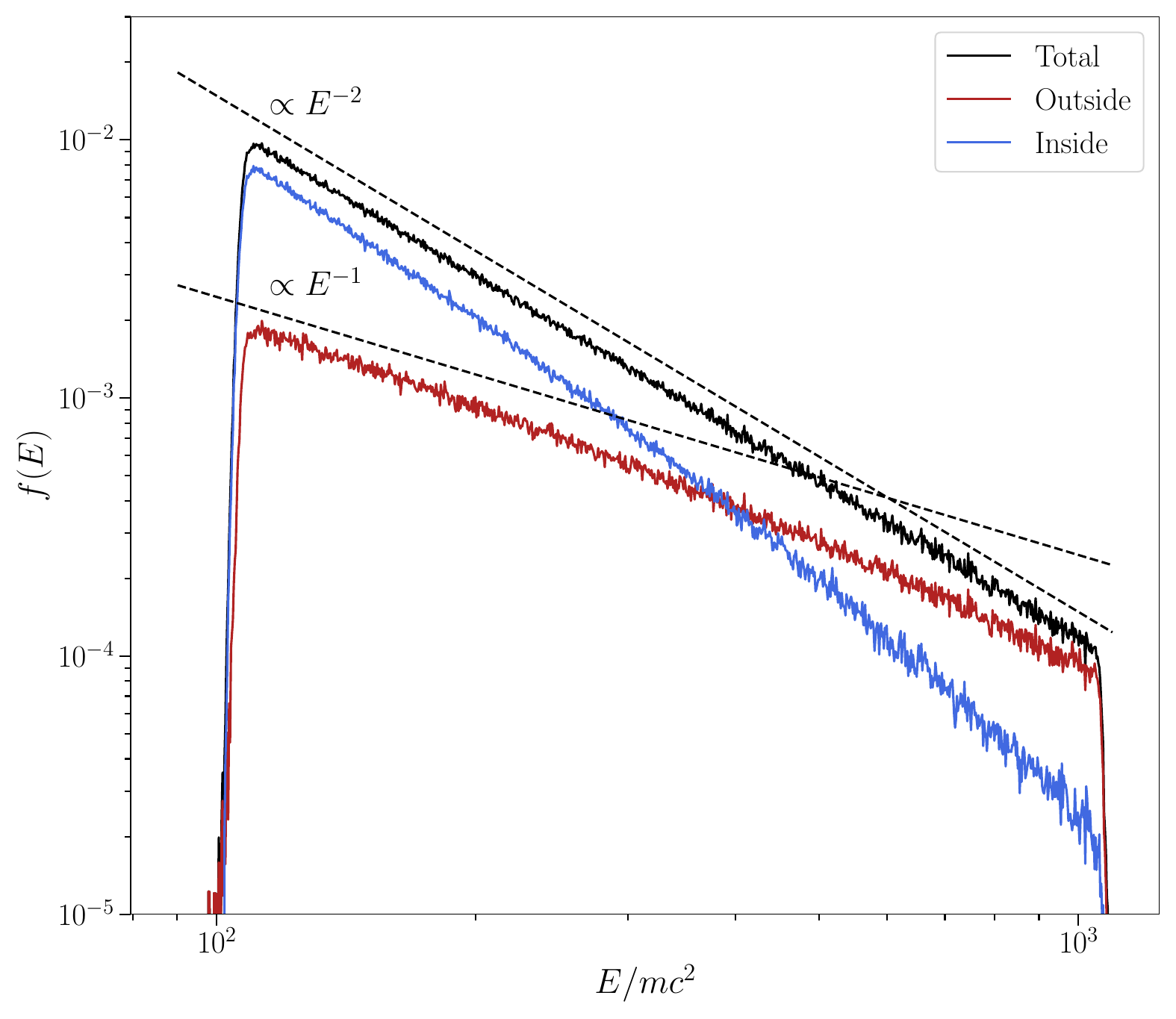}
\caption{The distribution function $f(E)$ of CR energies inside (blue) and outside (red) of the bubble after the bubble has risen~$0.5r_{b}$. The total distribution of all CRs is shown in black. For this simulation, CRs were sourced inside the bubble with a distribution of energies $\propto E^{-2}$. CRs of higher energies preferentially leak from the bubble, hardening the spectra outside the bubble in agreement with (\ref{eqn:S1:E12}).}
\label{Fig:Effusion}
\end{figure}

That~(\ref{eqn:S1:E7}) has an energy dependence provides an observationally intriguing possibility for the radio bubbles. Since the CRs of higher energies escape the bubble faster, this bears a certain resemblance to an effusive process: CRs with higher energies will be overrepresented outside the bubble.

To make this statement more quantitative, consider the following argument. From~(\ref{eqn:S1:E7}), we may assume that the distribution of particle energies within the bubble will have form
\begin{equation}
\label{eqn:S1:E11}
f(E) = f_{0}(E)\exp\left[-\alpha \sqrt{\frac{\kappa(E)}{u_{b}r_{b}}} \frac{u_{b}t}{r_{b}}\right],
\end{equation}
where $\alpha$ is some order-unity constant related to the precise nature of the flow. The energy dependence enters via the injected spectrum~$f_{0}(E)$ and the diffusion coefficient~$\kappa(E)$. For times earlier than the characteristic loss time (\ref{eqn:S1:E7p5}) of CRs at a given energy, the exponent in (\ref{eqn:S1:E11}) will be small. While the CR distribution inside the bubble will, therefore, change little, outside the bubble the distribution will~be
\begin{equation}
\label{eqn:S1:E12}
\begin{split}
f_{\mathrm{out}}(E) & = f_{0}(E)\left\lbrace 1 -  \exp\left[-\alpha \sqrt{\frac{\kappa(E)}{u_{b}r_{b}}} \frac{u_{b}t}{r_{b}}\right]\right\rbrace \\& \sim f_{0}(E)\alpha \sqrt{\frac{\kappa(E)}{u_{b}r_{b}}} \frac{u_{b}t}{r_{b}} \propto f_{0}(E) E,
\end{split}
\end{equation}
the last proportionality following from (\ref{eqn:S1:E5}). Thus, the CR spectrum will be hardened by a factor of $E$ outside the bubble. We confirm this result in Figure~\ref{Fig:Effusion}, which indeed shows the hardening of the CR spectra of CRs outside the bubble. This also offers an explanation for the hardening of the CR spectrum outside radio filaments recently reported in the giant fossil radio lobe of the Ophiuchus galaxy cluster (\citealt*{Giacintucci_2024}).

Another way of understanding this effect is as follows. The typical density of the layer of CRs around the bubble will be $\propto f_{0}(E)$ since the density of CRs must be continuous across the boundary. However, the boundary-layer width $\Delta r$ is a function of energy (\ref{eqn:S1:E6.5}), $\Delta r \propto E$, so the distribution of energies outside the bubble and in the wake will be proportional to the density imposed by the source multiplied by this width. 
\section{Conclusion and Discussion}
\label{Section:Conclusion}
We have proposed a new model for the confinement of CRs within radio bubbles, offering a possible explanation for the sharp boundaries seen in radio observation of these bubbles. While the model is simple in nature, being based straightforwardly on the competition between advection and diffusion of CRs, its success hinges on the enhanced CR scattering proposed by \cite{Reichherzer_2023}. We confirm numerically (see Figure \ref{Fig:numerical result}) that such an enhanced scattering does indeed provide good confinement of sub-TeV CRs and argue that the alternative, more conventional schemes for CR scattering at such energies are too weak to provide adequate CR confinement via isotropic diffusion. Finally, we note that the energy dependence of the CR diffusion coefficient has an interesting observational implication for the observed CR spectrum outside the bubble, which we predict (provided it has not been significantly aged by radiation losses) to be one power of energy shallower than the source spectrum (see Figure~\ref{Fig:Effusion}), due to the greater ease with which high-energy CRs can leave the bubble. 

A promising consequence of the long, energy dependent, confinement time (\ref{eqn:S1:E7p5}) of CRs is that it is on the same order as, and not in great excess of, the lower bound on confinement time inferred by \cite{Prokhorov_2017} from the lack of observed gamma-ray emission. While the two estimates should not be directly compared like for like (as their model assumed energy-independent escape and diffusion timescales), the proximity of the two estimates implies that diffusion of CRs in micromirrors could play an important role in constraining the energetic content of the bubbles (see, e.g., \citealt{Beckmann_2022};\citealt*{Yang_2019}; \citealt{Ruszkowski_2023}). 

A number of approximations have been made in order to arrive at these conclusions. One of our more extreme modelling assumptions is the artificial imposition of the bubble boundary and flow profile. While we anticipate that the true velocity field around the bubble will be different in detail from that assumed by us, it will undoubtedly contain a wake and a flow around the bubble, which are the only essential ingredients in our picture. This does, however, presuppose the integrity of the bubble to hydrodynamic instabilities, which would otherwise shred it. While the enhanced scattering offered by micromirrors can adequately confine CRs, those same micromirrors may also suppress the ICM viscosity (see, e.g., \citealt{Kunz_2014}; \citealt*{Melville_2016}). Should the ICM viscosity be overly suppressed, this would lead to a disruption of bubbles, which we clearly do not see (see discussion in, e.g., \citealt{Ruszkowski_2007,Kingsland_2019}). Of course, for our model to work, it is sufficient to have only the immediate vicinity of the bubble's boundary infested by micromirrors, since the CRs do not sample a large volume around the bubble. As discussed in Section \ref{Section:micromirror_diffusion_coefficient}, one could, therefore, envision a possibility where micromirrors are localised to the area of increasing magnetic-field strength around the bubble (where they are naturally driven unstable by the resulting pressure anisotropy), while the ICM at large may have a much smaller fraction of its volume filled by micromirrors. Furthermore, the same pressure anisotropy that creates the micromirrors also enhances the effective tension force exerted by the magnetic draping, potentially protecting the bubble from hydrodynamic instabilities in spite of the lower viscosity.

We also do not capture the effects of bubble deformation (\citealt{Guo_2015}) and the back reaction of the CRs on the bubble. Should such back reaction be included, it is now clear that it must also include the enhanced diffusion of CRs by micromirrors, if for no other reason than that the dimensional estimate (\ref{eqn:S1:E5}) ensures its \textit{a priori} non-negligibility. As for the effects of bubble growth and of the variation in the flow speed, our model suggests the advantageous picture that the bubble will actually become \textit{better} at confining CRs as it rises, because one expects the bubbles to slow down and grow, increasing the confinement time~(\ref{eqn:S1:E7p5}),~$t_{\mathrm{conf}} \propto u_{b}^{-1/2}r_{b}^{3/2}$. This perhaps points to the possibility that CR scattering, especially by micromirrors, as a confinement mechanism could be effective for a wide range of ICM radio features that account for the earlier or later stages of the bubble's life, such as filamentary structures, lobes or jets. 

\section*{Acknowledgements}
It is a pleasure to thank Georgia Acton, Eugene Churazov, Plamen Ivanov, Philipp Kempski, Hrushikesh Loya, Michael Nastac, Marcel Rod, Luis Silva, and Dmitri Uzdensky for illuminating discussions. RJE was supported by a UK EPSRC studentship, PR by a Gateway Fellowship and Walter-Benjamin Fellowship, AFAB by a UKRI Future Leaders Fellowship (grant number MR/W006723/1), MWK in
part by NSF CAREER Award No. 1944972; the work of AAS was supported in part by grants from STFC (ST/W000903/1) and EPSRC (EP/R034737/1), as well as by the Simons Foundations via a Simons Investigator award.

\section*{Data availability}
The data underlying this article will be shared on reasonable request to the corresponding author.

\bibliographystyle{mnras}
\bibliography{Bubblebib} 




\appendix

\section{Numerical details of mock simulation}
\label{App:Numerical_details}
In this Appendix, we detail the numerical scheme by which the bubble and CRs were evolved in the mock numerical simulations whose results we presented in Figures \ref{Fig:numerical result}-\ref{Fig:Effusion}. We use a Monte Carlo method, initialising a large number of CRs uniformly inside the boundary of the bubble. There are then two principal components in the simulation: the specification of the fluid flow in the rest frame of the bubble and the evolution equations for the CRs. We discuss the two separately, in turn.
\subsection{Fluid flow around the bubble}
As illustrated in Figure \ref{Fig:Schematic}, the fluid flow that we impose is a 3D potential flow around a hemispherical cap representing the bubble, with an entrained wake represented by a column of fluid behind the bubble. It is known that the flow around a spherical object is well approximated by an incompressible, steady, irrotational flow. However, behind the bubble, such a flow would not capture the turbulent wake that is formed. To model this turbulent wake in the most brutal fashion possible, we continue the streamlines vertically downwards from the incompressible, steady, irrotational flow imposed above the sphere. This generates an incompressible rotational flow that describes the bubble dragging a column of fluid upwards from beneath it. The flow field in cylindrical coordinates $z,r,\varphi$ will have azimuthal symmetry and be given by the vector field $\v{u}(z, r) = u_{z}(z,r)\v{e}_{z} + u_{r}(z, r)\v{e}_{r}$, where the vertical velocity is
\begin{equation}
\label{eqn:A1:E1}
u_{z}(z,r) = \begin{cases} -v_{b}\left[1 -  r_{b}^{3}\displaystyle\frac{2z^{2} - r^{2}}{2\left(z^{2} + r^{2}\right)^{5/2}}\right] \, &\text{for} \, z>0, z^{2}+r^{2} > r_{b}^{2},
\\[10pt]  -v_{b}\left(1 + \displaystyle\frac{r_{b}^{3}}{2r^{3}}\right)  &\text{for} \, z<0, r > r_{b},
\\[10pt] 0  & \text{otherwise},
\end{cases}
\end{equation}
and the radial velocity is
\begin{equation}
\label{eqn:A1:E2}
u_{r}(z,r) = \begin{cases}
\displaystyle v_{b}r_{b}^{3}\frac{3z r}{2\left(z^{2} + r^{2} \right)^{5/2}} \quad &\text{for} \quad z>0, z^{2}+r^{2} > r_{b}^{2},
\\[10pt] 0 \quad &\text{otherwise}.
\end{cases}
\end{equation}
\subsection{Cosmic-ray evolution}
To model CR propagation, we associate to each CR initially a position~$\v{x}$, an orientation~$\v{n}$, and a gamma factor~$\gamma$. Then the question is how to step the position, orientation and gamma factor forward in time. We know from (\ref{eqn:S1:E5}) that the diffusion coefficient should be a function of the CR energy~$E = \gamma mc^{2}$, and, therefore, of $\gamma$. As the simplest model of diffusive scattering possible, we consider the case where the CRs propagate for a time $1/\nu(\gamma)$, whereupon, if they find themselves outside the bubble, they are isotropically scattered \textit{in the rest frame of the flow}. Explicitly this means that if~$\v{x}(t)$,~$\v{n}(t)$, and~$\gamma(t)$ are the position, orientation, and gamma factor of a CR at time~$t$, then its position at time~$t + \Delta t$ is simply 
\begin{equation}
\v{x}(t+\Delta t) = \v{x}(t) + c\Delta t \frac{\sqrt{\gamma(t)^{2} - 1}}{\gamma(t)}\v{n}(t),
\end{equation}
where $\Delta t = 1/\nu(\gamma(t))$. To find $\gamma(t)$ at the next time step, we first compute its value $\tilde{\gamma}$ in the frame moving with the local flow~{$\v{u}(\v{x}(t+\Delta t))$:}
\begin{equation}
\tilde{\gamma}(t) = \gamma_{\v{u}}(\v{x}(t + \Delta t))\left[\gamma(t)  - \sqrt{\gamma(t)^{2} - 1}\frac{\v{n}(t)\cdot\v{u}(\v{x}(t+\Delta t))}{c} \right],
\end{equation}
where $\gamma_{\v{u}}$ is the gamma factor of the local flow, evaluated at~{$\v{x}(t + \Delta t)$}---very close to unity,  but included for completeness. The CR will scatter to a random orientation, denoted $\tilde{\v{n}}$---isotopic in the rest frame of the flow. Boosting back into the laboratory frame, we use the scattered orientation $\tilde{\v{n}}$ to find the value of the gamma factor at the next time step:
\begin{equation}
\gamma(t+\Delta t) = \gamma_{\v{u}}(\v{x}(t+ \Delta t))\Big[\tilde{\gamma}(t) + \sqrt{\tilde{\gamma}(t)^{2} - 1}\frac{\tilde{\v{n}}\cdot\v{u}(\v{x}(t + \Delta t))}{c} \Big],
\end{equation}
and the scattered orientation in the laboratory frame:
\begin{equation}
\begin{split}
\v{n}(t + \Delta t) = & \frac{\sqrt{\tilde{\gamma}(t)^{2} - 1}}{\sqrt{\gamma(t+\Delta t)^{2} - 1}}\Bigg\lbrace\tilde{\v{n}} + \\ & +  \left[\gamma_{\v{u}}(\v{x}(t+ \Delta t)) - 1\right]\frac{\tilde{\v{n}}\cdot\v{u}(\v{x}(t + \Delta t))}{|\v{u}(\v{x}(t + \Delta t))|}\frac{\v{u}(\v{x}(t + \Delta t))}{|\v{u}(\v{x}(t + \Delta t))|} \Bigg\rbrace \\ & + \frac{\tilde{\gamma}(t)\gamma_{\v{u}}}{\sqrt{\gamma(t+\Delta t)^{2} - 1}}\frac{\v{u}(\v{x}(t + \Delta t))}{c},
\end{split}
\end{equation}
which naturally satisfies $|\v{n}| = 1$ at each time step.
\section{Energisation of Cosmic rays}
\label{App:Energy}
\begin{figure*}
\centering
\includegraphics[scale=0.48]{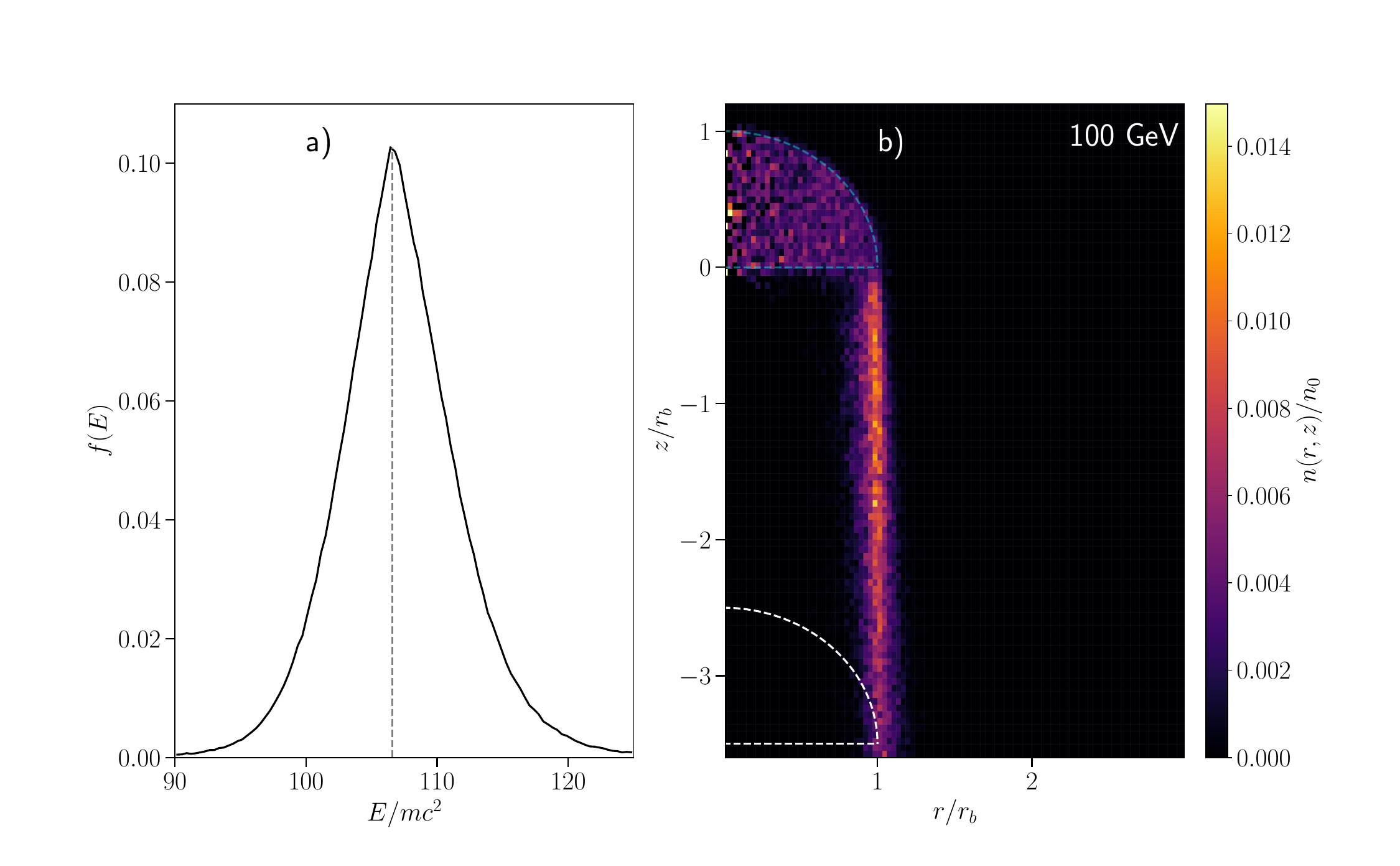}
\caption{(a) Numerically obtained distribution function, $f(E)$, of CRs (solid line) achieved after the bubble has risen $3.5r_{b}$. The CRs were initialised at $100\, \mathrm{GeV}$ (dashed line). (b) Density $n_{10\%}$ of CRs (as in Figure\ref{Fig:numerical result}) filtered to highlight only those CRs that have increased their energy by more than $10\%$.}
\label{Fig:Energisation}
\end{figure*}
While the numerical method presented in Appendix \ref{App:Numerical_details} allows for the variation in the energy of the CRs due to \cite{Fermi_1949} acceleration (i.e., energy gain via scattering off moving parcels of fluid), this process is ignored in our theoretical considerations of the boundary layer thickness and CR confinement time in Section \ref{Section:central_concept}. In this appendix, we justify this omission, showing that energisation of CRs due to Fermi acceleration is at most an order-unity effect, and typically negligible.

Figure \ref{Fig:Energisation} (a) shows the energy distribution after the bubble has risen $3.5r_{b}$ of a population of CRs initialised at $100\, \mathrm{GeV}$. We see, by eye, that the change of energy is largely diffusive, and that it is typically on the order of $\sim 5\%$. That the energy change should be diffusive is expected for Fermi acceleration (see, e.g., \citealt{Lemoine_2019} and references therein). Furthermore, Figure \ref{Fig:Energisation} (b) tells us that the CRs that have diffused most in energy are found in the population escaped from the bubble. This is a simple matter to explain, as the CRs that have escaped from the bubble are likely to have undergone many scatterings on its surface, diffusing the furthest in energy.

To make this statement more quantitative, we note that the strongest kicks in Fermi acceleration should come from the CRs crossing from the bubble interior, where the flow is stationary (in the rest frame of the bubble), to the flow outside the bubble, with speed~$\sim\! u_{b}$, experiencing a relative change in energy $\sim u_{b}/c$. The typical CR will make many such crossings of the bubble boundary, each time being recaptured, until it eventually escapes after a time $t_{\mathrm{conf}}$ given by~(\ref{eqn:S1:E7p5}). Since it takes a time~$\sim\! r_{b}/c$ to cross the bubble, the number of crossings of the bubble boundary that a typical CR will make before escaping is
\begin{equation}
N_{\mathrm{esc}} \sim \frac{c}{u_{b}}\sqrt{\frac{u_{b}r_{b}}{\kappa}}.
\end{equation}
Since changes of the CR energy made in each crossing add up as a random walk, the typical energy change before escape will be 
\begin{equation}
\begin{split}
\frac{\Delta E}{E} &\sim \frac{u_{b}}{c}\sqrt{N_{\mathrm{esc}}} \sim \sqrt{\frac{u_{b}}{c}}  \left(\frac{u_{b}r_{b}}{\kappa}\right)^{1/4} \\ & \sim 0.03\left(\frac{E}{\mathrm{TeV}}\right)^{-1/2}\left(\frac{u_{b}}{200\,\mathrm{km}\,\mathrm{s^{-1}}} \right)^{3/4}\left(\frac{r_{b}}{10\,\mathrm{kpc}} \right)^{1/4}.
\end{split}
\end{equation} 
This is in rough agreement with the $\sim 10-20\%$ acceleration/deceleration for $100\, \mathrm{GeV}$ CRs seen in Figure \ref{Fig:Energisation}. This tells us that we can expect order-unity energy changes for the lowest-energy CRs (those of a few $\mathrm{GeV}$). 
\section{Asymptotic solution of Blasius problem above bubble}
\label{App:Asym}
\begin{figure*}
\centering
\includegraphics[scale=0.45]{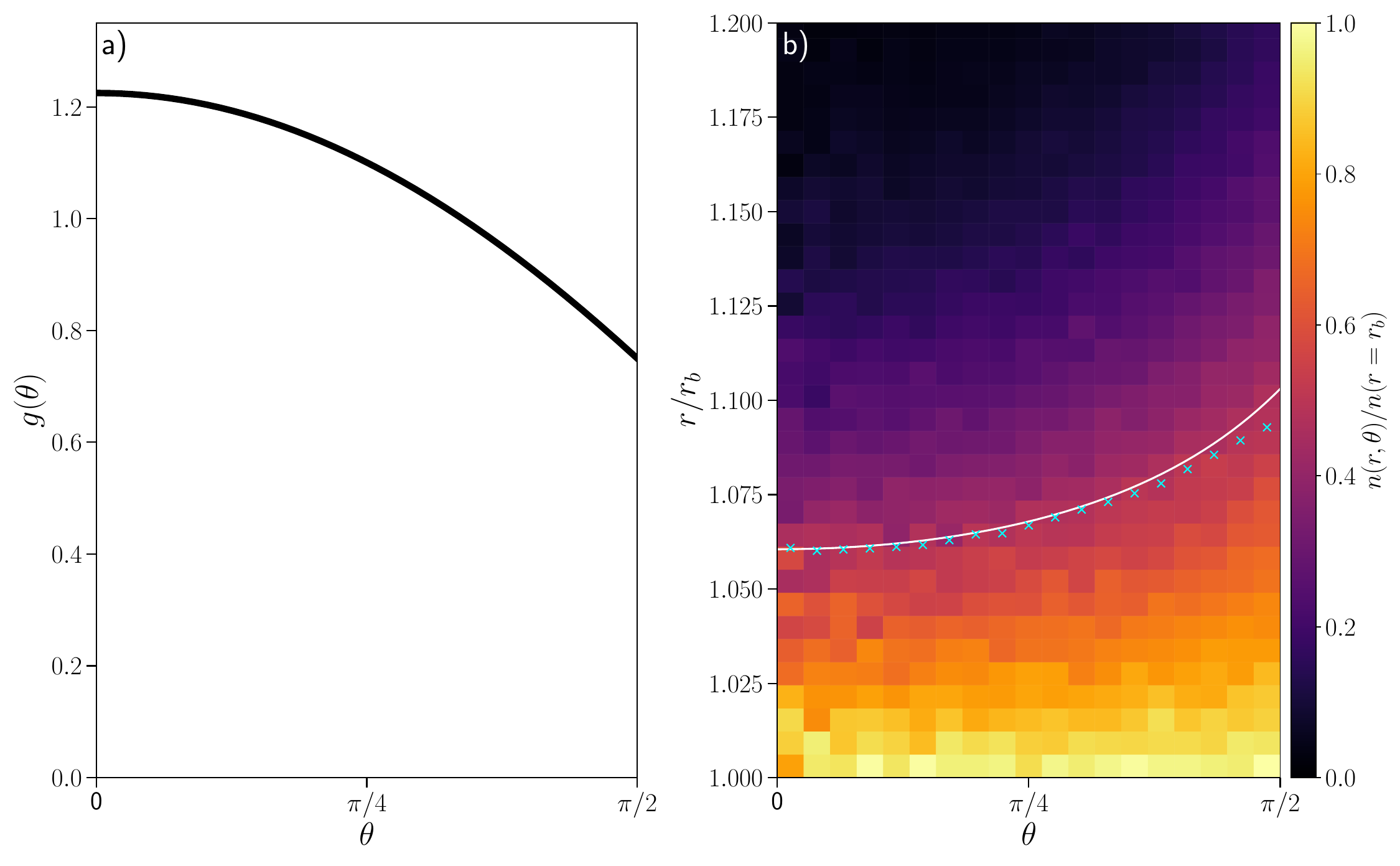}
\caption{(a) Numerical solution of (\ref{eqn:A2:E6}), giving the function $g(\theta)$ that controls the width of the boundary layer around the top of the bubble. (b) Numerical simulation of the density of CRs at $300\,\mathrm{GeV}$ outside the bubble assuming volume-filling micromirrors. Crosses show the mean radial distance of CRs from the nominal bubble boundary in each angular interval, the solid line shows the mean radial distance of CRs from the bubble (\ref{eqn:A2:E9}) calculated from (\ref{eqn:A2:E8}) with $g(\theta)$ taken from panel (a).}
\label{Fig:Blasius_Compare}
\end{figure*}
In this appendix, we solve the Blasius problem in the limit
\begin{equation}
\label{eqn:A2:E1}
\epsilon \equiv \frac{\kappa}{u_{b}r_{b}} \ll 1.
\end{equation}
The Blasius problem has previously been treated by \cite{Churazov_2004} and \cite{Xiang_2007} in determining the width of cold fronts due to thermal conduction in a potential flow past a sphere. They solve, analytically, for the steady state of the advection-diffusion equation near the leading edge of the cold front. Their solution confirms that the width of the layer produced in front of the bubble is of the order $\sqrt{\varepsilon}r_{b}$. Here, we are principally interested in the scaling of the width of the layer at the side of the bubble, since it is this layer that determines the rate of particle loss from the bubble. To determine this width, we show that a solution with a layer of width~$\sqrt{\varepsilon}r_{b}$ exists for all polar angles $\theta \leq \pi/ 2$, extending the solution of \cite{Xiang_2007} away from the top ($\theta = 0$) of the bubble. 

In spherical geometry and with the flow specified by~(\ref{eqn:A1:E1}) and~(\ref{eqn:A1:E2}), the steady-state advection-diffusion equation (\ref{eqn:S1:E6}) above the bubble~($\theta \leq \pi/ 2$) becomes 
\begin{multline}
\label{eqn:A2:E2}
-u_{b}\left(1 - \frac{r_{b}^{3}}{r^{3}} \right)\cos\theta \pdev{n}{r} + u_{b}\left(1  + \frac{r_{b}^{3}}{2r^{3}}\right)\sin\theta \frac{1}{r}\pdev{n}{\theta}\\ - \frac{\kappa}{r^{2}}\left(\pdev{}{r}r^{2}\pdev{n}{r}  + \frac{1}{\sin\theta}\pdev{}{\theta}\sin\theta\pdev{n}{\theta} \right) = 0.
\end{multline}
To show that this equation admits a solution with a layer of thickness~$\sqrt{\varepsilon}r_{b}$, we make the change of variables~{$r = r_{b}(1 + \sqrt{\varepsilon} x)$} and retain only the lowest-order terms: 
\begin{equation}
\label{eqn:A2:E3}
-3x\cos\theta \pdev{n}{x} + \frac{3}{2}\sin\theta\pdev{n}{\theta} - \pdevn{n}{x}{2} = 0.
\end{equation}
Note that, at this order, we have thrown away the polar-diffusion term---the $4^{\mathrm{th}}$ term in (\ref{eqn:A2:E2}), which can break the ordering at small~$\theta$. This is fine, however, provided we assume $\theta \gg \sqrt{\varepsilon}$ and find a solution that satisfies
\begin{equation}
\label{eqn:A2:E4}
\left.\pdev{n}{\theta}\right|_{\theta \to 0} = 0.
\end{equation}

We now seek a solution to (\ref{eqn:A2:E3}) in the form
\begin{equation}
\label{eqn:A2:E5}
n(x,\theta) = n_0\left\lbrace 1 - \mathrm{erf}\left[g(\theta)x\right] \right\rbrace.
\end{equation}
This is indeed a \textit{bona fide} solution provided $g(\theta)$ satisfies the differential equation
\begin{equation}
\label{eqn:A2:E6}
\sin \theta \pdev{g}{\theta} = 2\cos\theta g(\theta) - \frac{4}{3}g^{3}(\theta).
\end{equation}
To enforce (\ref{eqn:A2:E4}) without breaking the ordering of $\partial n / \partial x = \mathrm{O}\!\left(1\right)$, we must, therefore, have 
\begin{equation}
\label{eqn:A2:E7}
g(\theta  \to 0) = \sqrt{\frac{3}{2}}.
\end{equation}
Thus, the lowest-order solution above the bubble is given by
\begin{equation}
\label{eqn:A2:E8}
n(r, \theta, \gamma) = n_{0}\left\lbrace 1 - \mathrm{erf}\left[\sqrt{\frac{u_{b}r_{b}}{\kappa}}g(\theta)\frac{r - r_{b}}{r_{b}} \right] \right\rbrace.
\end{equation}
Since $g(\theta)$ is finite between $0\leq \theta \leq \pi/2$, this ensures that the ordering (\ref{eqn:S1:E6.5}) is valid everywhere above the bubble. The numerically integrated solution of (\ref{eqn:A2:E6}) for $g(\theta)$ is shown in Figure~\ref{Fig:Blasius_Compare} (a). In Figure \ref{Fig:Blasius_Compare} (b), the mean radial distance of CRs from the bubble
\begin{equation}
\label{eqn:A2:E9}
\crl{r}(\theta) = \frac{\int_{r_{b}}^{\infty} r^{3} n(r,\theta)\dd{r}}{\int_{r_{b}}^{\infty} r^{2} n(r)\dd{r}},
\end{equation}
is compared, favourably, to the result of our mock simulation above the bubble.  \addcom{FUN FACT: you can also solve this same problem for a creeping (Stokes) flow past a sphere, and in that case the boundary width scale as $\varepsilon^{1/3}r_{b}$: viscous bubbles have thicker skins. But they would still lose CRs slower because of their no slip boundary.}

\bsp	
\label{lastpage}
\end{document}